\documentclass[aps,pra,twocolumn,superscriptaddress]{revtex4}

\usepackage{amsmath,amsfonts,amssymb}
\usepackage{graphicx,float,calc}
\usepackage{color,bm}
\usepackage{ulem}
\usepackage{braket}
\usepackage{hyperref}
\usepackage{graphicx,epstopdf}
\usepackage{makecell}
\usepackage{multirow}
\usepackage{array}
\usepackage{CJK}
\usepackage{txfonts}

\newcommand{\ehbar}{\hbar_{\mathrm{eff}}}

\makeatother

\begin{document}
\title{Super-exponential behaviors of out-of-time ordered correlators and Loschmidt echo in a non-Hermitian interacting system}

\author{Wen-Lei Zhao}
\email{wlzhao@jxust.edu.cn}
\affiliation{School of Science, Jiangxi University of Science and Technology, Ganzhou 341000, China}
\affiliation{Key Laboratory of Low Dimensional Quantum Materials and Sensor Devices of Jiangxi Education Institutes, Ganzhou 341000, China}

\author{Chao Han}
\affiliation{School of Science, Jiangxi University of Science and Technology, Ganzhou 341000, China}
\affiliation{Key Laboratory of Low Dimensional Quantum Materials and Sensor Devices of Jiangxi Education Institutes, Ganzhou 341000, China}

\author{Han Ke}
\affiliation{School of Science, Jiangxi University of Science and Technology, Ganzhou 341000, China}

\author{Jie Liu}
\email{jliu@gscaep.ac.cn}
\affiliation{Graduate School, China Academy of Engineering Physics, Beijing 100193, China}
\affiliation{HEDPS, Center for Applied Physics and Technology, and College of Engineering, Peking University, Beijing 100871, China}

\begin{abstract}
We investigate the out-of-time ordered correlators and Loschmidt echo in a non-Hermitian interacting system governed by a Gross-Pitaevskii map model, which incorporates a periodically modulated complex strength of the nonlinear interaction as delta kicks. We uncover that the time evolutions of the out-of-time ordered correlators follow that of the Loschmidt echo in certain situations. In particular, we find that both of them can exhibit a super-exponential growth with time, indicating the emergence of super-exponential scrambling and instability. Interestingly, after a proper scaling scheme, we find that all the super-exponential behaviors approximately collapse on a scaling-law curve that is independent on the non-Hermitian parameter as well as the effective Planck constant. The underlying mechanism is rooted in the super-exponentially fast diffusion of energy as well as the norm of quantum states. Our findings suggest a kind of fastest divergence of two nearby quantum states, which has implication in information scrambling.
\end{abstract}
\date{\today}

\maketitle

\section{Introduction}
How quickly can two nearby states diverge from one another in a quantum unstable system? This fundamental question consistently garners intensive attention across various fields of physics, including quantum chaos~\cite{WWang07}, quantum computation~\cite{Georgeot01}, and high energy physics~\cite{Qu22a,Qu22b}. The reason lies in its critical implications for both the theoretical prediction of quantum instability, as quantified by the Loschmidt Echo (LE)~\cite{Peres,Gorin06}, and potential applications in quantum scrambling signaling via out-of-time ordered correlators (OTOCs)~\cite{Harris2022,Zhang2019,Yan2020,Patel2017,Rozenbaum17,Maldacena16,JWang20,JWang21,Das21,Caputa17,BYan20,Sreeram21}.
The dynamical instability has a crucial bearing on the irreversibility of dynamical evolution, thus controlling the ``arrow of time''~\cite{Eddington28}. The time irreversibility has been used to detect quantum decoherence~\cite{Jalabert01} and quantum-classical transition~\cite{Benenti02,PQin12}, with its singularity represented in the LE employed to diagnose quantum phase transitions in condensed matter physics~\cite{Quan06,QWang15,PQin14,WWang12}. Intrinsically, OTOCs serve as valuable tools for quantifying quantum entanglement~\cite{Lerose20,ZQi23} and non-equilibrium dynamics~\cite{Balachandran21,SZhao22}. In quantum many-body systems, the OTOCs' dynamics can detect the effects of both interaction~\cite{Roberts2022} and interference~\cite{Rammensee18} on quantum thermalization and time reversibility. For a Gross-Pitaevskii map system, the temporally-modulated nonlinear interaction even induces super-exponential growth of OTOCs, highlighting the existence of a novel chaotic dynamics~\cite{WLZhao21}.

Despite fascinating advancements in Hermitian systems, quantum scrambling in non-Hermitian systems remains an intriguing yet elusive issue. Notably, the dynamics of OTOCs encircling exponential points (EPs) of a non-Hermitian Ising chain is governed by the Yang-Lee edge singularity~\cite{LZhai20}. In a $\cal{PT}$-symmetric chaotic system, the OTOCs even exhibit quantized response to external driven potentials~\cite{Zhao22} and scaling-laws at the transition to EPs~\cite{Zhao23arx,WLZhao23FIP}. Non-Hermitian physics has proven to be essential in various applications, such as enhancing the accuracy of quantum sensors~\cite{Chu20,Heiss12,Wiersig16,Liu16}, manipulating nonreciprocal light propagation~\cite{Yin13}, and engineering topological energy transfer in cryogenic optomechanical devices~\cite{Xu16}. The investigations on non-Hermitian physics have given rise to important breakthroughs in fundamental concepts~\cite{Hatano96PPL,Pancharatnam,Berry94,Berry04}, including the spontaneous breaking of chiral symmetry~\cite{Stephanov96,Markum99}, the non-equilibrium dynamics~\cite{Kadanoff68}, as well as the non-Hermitian topology~\cite{Bergholtz21,Yao18,Lee19,LLi20,Okuma20,Borgnia20,Song19,Jiang19}.

There are still many open questions arising from the fundamental concept of quantum chaos with non-Hermiticity and inter-particle interaction. For example, are there quantum systems that display an even higher degree of instability than the traditional exponentially instability? In this work, we investigate both the information scrambling and dynamical instability in a non-Hermitian Gross-Pitaevskii map (NGPM) model with delta-kicking modulation of complex nonlinear interaction. We demonstrate a rigorous equivalence between the fidelity of two quantum states and OTOCs by properly selecting the appropriate operators for its construction. Furthermore, we demonstrate that the mean energy increases with time in a super-exponential function, which gives rise to both the super-exponential growth of OTOCs and the distance between two nearby quantum states.
After a proper scaling scheme, we find that all the super-exponential behaviors approximately collapse on a scaling-law curve that is independent on the non-Hermitian parameter as well as the effective Planck constant. The phenomena of both super-exponential instability and super-exponential scrambling are expected to serve as new elements in the fields of quantum chaos and quantum information with regard to non-Hermitian and nonlinear systems.

\section{Model and main results}

The Hamiltonian of a NGPM reads
\begin{equation}\label{NGPHamil}
\text{H}= \frac{p^2}{2} + (g+i\eta) |\psi(\theta)|^2\sum_n \delta (t-t_n)\;,
\end{equation}
where $p=-i\ehbar\partial/\partial \theta$ is the angular momentum operator, $\theta$ is the angle coordinate, with the commutation relation $[\theta, p]=i\ehbar$ and $\ehbar$ indicating the effective Planck constant. The parameters $g$ and $\eta$ control the strength of the real and imaginary parts of the nonlinear interaction, respectively. The time $t_n$ denotes kick number thus is integer, i.e., $t_n=0,1,2\ldots$. All variables are properly scaled, hence in dimensionless units. The time evolution from $t=t_n$ to $t=t_{n+1}$ is governed by $|\psi(t_{n+1})\rangle=U|\psi(t_{n})\rangle$, where the Floquet operator takes the form
\begin{equation}\label{Evoopt}
U(t_{n})= U_f U_K(t_n)
\end{equation}
with free evolution operator
\begin{equation}\label{FreEvoopt}
U_f= \exp\left(-\frac{i}{\ehbar}\frac{p^2}{2}\right)\;,
\end{equation}
and kicking evolution operator
\begin{equation}\label{KikEvoopt}
U_K(t_{n})= \exp\left[\frac{-i(g+i\eta)|\psi(\theta,t_n)|^2}{\ehbar}\right]\;.
\end{equation}

The OTOCs are defined as  $C=-\langle [W(t),V]^2\rangle$, where the operators $W(t)=U^{\dagger}(t)WU(t)$ and $V$ are evaluated in the Heisenberg picture, and $\langle \cdots\rangle = \langle \psi(t_0)|\cdots|\psi(t_0)\rangle$ indicates an average over the initial state. Using the translation operator $W=e^{-i\varepsilon p/\ehbar}$ and a density operator of an initial state $V=\rho(t_0)=|\psi(t_0)\rangle\langle\psi(t_0)|$, one obtains the relation relation $C(t)=\mathcal{N}^2_{\psi}(t)-|\langle \psi(t)|e^{-i\varepsilon p/\ehbar}|\psi(t)\rangle|^2$~\cite{WLZhao24PRA,WLZhao24PRR}, where $\mathcal{N}_{\psi}(t)=\langle \psi(t)|\psi(t)\rangle$ denotes the norm of quantum state. For non-Hermitian systems, the non-unitary evolution results in the increase of the norm with time. Therefore, a natural definition of the rescaled OTOCs is given by
\begin{equation}\label{RSOTC}
C(t)=1-|\langle \psi(t)|e^{-i\varepsilon p/\ehbar}|\psi(t)\rangle|^2/\mathcal{N}^2_{\psi}(t)\;,
\end{equation}
which eliminates the contribution of the norm~\cite{WLZhao24PRA,WLZhao24PRR}. Note that the term $|\langle \psi(t)|e^{-i\varepsilon p/\ehbar}|\psi(t)\rangle|^2$ is usually referred to as Fidelity OTOCs (FOTOC)~\cite{Lewis-Swan19}. A straightforward definition of the distance between two quantum states is given by
\begin{equation}\label{Distance}
D(t) = 1 - \frac{|\langle \psi(t) | \varphi(t) \rangle|^2}{\mathcal{N}_{\psi}(t) \mathcal{N}_{\varphi}(t)}\;,
\end{equation}
where $\mathcal{N}_{\psi}$ and $\mathcal{N}_{\varphi}$ denote the norm of the respective states.
Under the condition that the two initial states are related by
$|\varphi(t_0)\rangle = T(-\varepsilon) |\psi(t_0)\rangle$,
we establish a strict equivalence between $C(t)$ and the distance $D(t)$, namely,
\begin{equation}\label{SExpDist}
C(t) = D(t)\;.
\end{equation}

\begin{figure}[t]
\begin{center}
\includegraphics[width=7.5cm]{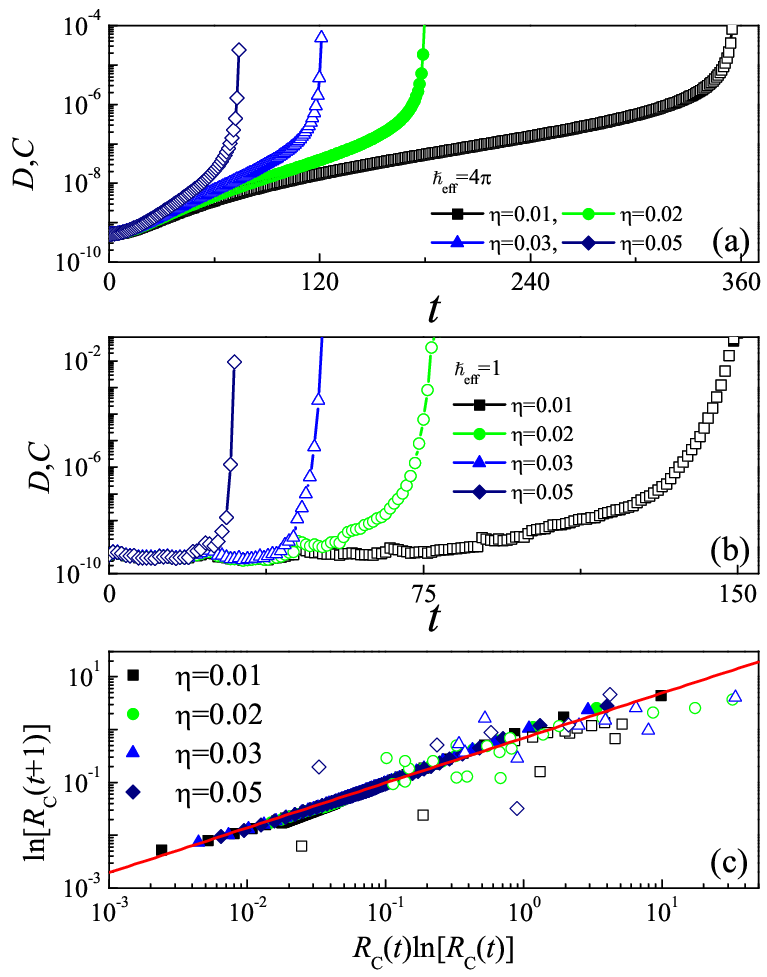}
\caption{(a) and (b): Time dependence of $C$ (solid symbols) and $D$ (open symbols), in semi-log scale, for $\ehbar=4\pi$ (a) and 1 (b), with $\eta=0.01$ (squares), $0.02$ (circles), 0.03 (triangles), and 0.05 (diamonds). (c) The $\ln[R_{C}(t+1)]$ versus $R_{C}(t)\ln[R_{C}(t)]$, in log-log scale, for $\ehbar=4\pi$ (solid symbols) and 1 (open symbols), with $\eta=0.01$ (squares), $0.02$ (circles), 0.03 (triangles), and 0.05 (diamonds). Solid line indicates the fitting function $\ln[R_{C}(t+1)]=\alpha [R_{C}(t)\ln[R_{C}(t)]]^{\beta}$, where $\alpha \approx 0.69$ and $\beta \approx 0.85$. The nonlinear interaction strength is $g=0.3$.}\label{DisOTOCs}
\end{center}
\end{figure}

We numerically investigate the time evolution of both the $C$ and $D$ for different $\ehbar$. We assume, without loss of generality, that the initial state $|\psi(t_0)\rangle$ is a Gaussian wavepacket, defined as $\psi{(\theta,t_0)}=(\sigma/\pi)^{1/4} \exp (-\sigma \theta^{2}/2)$ with $\sigma=10$. Another quantum state $|\varphi(t_0)\rangle$ is defined as $\varphi(\theta,t_0)=T(-\varepsilon)\psi{(\theta,t_0)}=\psi{(\theta+\varepsilon,t_0)}$ with $\varepsilon=10^{-5}$.
We first focus on the quantum resonance condition, i.e., $\ehbar = 4\pi$, under which the free evolution operator $U_f$ is the identity, as $U_f(p_n) = \exp\left(-i n^2 2\pi\right) = 1$. Figure~\ref{DisOTOCs}(a) shows the perfect agreement between $C$ and $D$ for different $\eta$, confirming our theoretical prediction in Eq.~\eqref{SExpDist}. Moreover, for nonzero $\eta$, they increase super-exponentially with time, and the growth rate increases as $\eta$ increases. We further consider the non-resonant case, i.e., $\ehbar = 1$, and numerically find the equivalence $C = D$ as well as their superexponentially fast increase. Note that the exponential increase of $D$ mimics the super-exponential divergence of two nearby trajectories~\cite{WLZhao16,WLZhao19jpa,Guarneri17}. Accordingly, our system exhibits super-exponential instability arising from the interplay between non-Hermiticity and nonlinearity, a feature absent in conventional chaotic systems. Since the FOTOC is a common measure of quantum scrambling~\cite{Lewis-Swan19}, the mathematical equivalence $C = D$ implies that the underlying instability drives the super-exponential scrambling of quantum information.

To uncover the governing law of this super-exponential behavior, we introduce the ratio of OTOCs between successive kicks as $R_C(t)=C(t+1)/C(t)$. Our numerical analysis reveals that, in the quantum resonance case, the ratio follows the relation
$\ln[R_{C}(t+1)]=\alpha [R_{C}(t)\ln[R_{C}(t)]]^{\beta}$
for different values of \( \eta \), with coefficients $\alpha\approx 0.69$ and $\beta\approx 0.85$. This demonstrates a scaling law behind the super-exponential behaviors of information scrambling. Under quantum non-resonance condition, the relation between $\ln[R_{C}(t+1)]$ and $R_{C}(t)\ln[R_{C}(t)]$ exhibit some fluctuations around the scaling law. It is known that, under the quantum non-resonance condition, the free evolution operator in the momentum representation effectively generates pseudorandom variables, which act as disorder, leading to dynamical localization in momentum space~\cite{Grempel84pra}. Thus, it is reasonable to attribute certain fluctuations in the scaling law to the effects of momentum-space disorder.

\begin{figure}[t]
\begin{center}
\includegraphics[width=8.0cm]{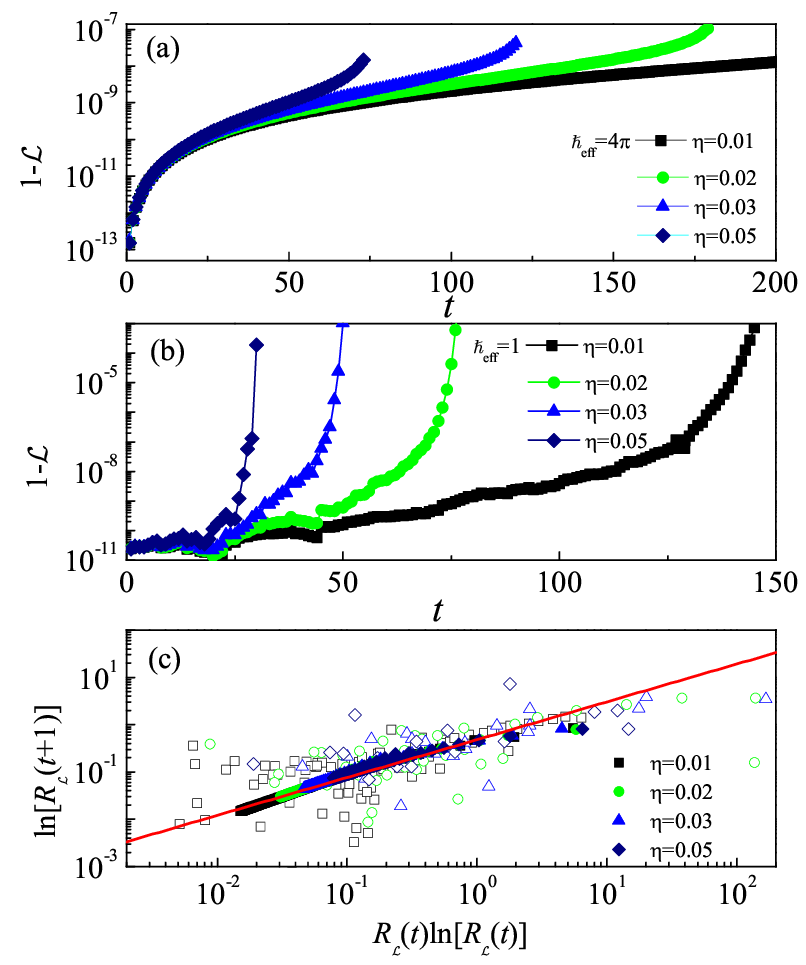}
\caption{(a) and (b): Time dependence of $1-\cal{L}$,  in semi-log scale, for $\ehbar=4\pi$ (a) and 1 (b), with $\eta=0.01$ (squares), $0.02$ (circles), 0.03 (triangles), and 0.05 (diamonds). (c) The $\ln[R_{\mathcal{L}}(t+1)]$ versus $R_{\mathcal{L}}(t)\ln[R_{\mathcal{L}}(t)]$, in log-log scale, for $\ehbar=4\pi$ (solid symbols) and 1 (empty symbols), with $\eta=0.01$ (squares), $0.02$ (circles), 0.03 (triangles), and 0.05 (diamonds). Solid line represents the fitting function $\ln[R_{\mathcal{L}}(t+1)]=\alpha [R_{\mathcal{L}}(t)\ln[R_{\mathcal{L}}(t)]]^{\beta}$, with $\alpha \approx 0.48$ and $\beta \approx 0.8$. The value of the translation parameter is $\varepsilon=10^{-5}$, and other parameters are the same as in Fig.~\ref{DisOTOCs}(a).}\label{LoschtEcho}
\end{center}
\end{figure}

In quantum chaos, the extreme sensitivity of wavepacket's dynamics to perturbations in the Hamiltonian is usually quantified by the LE $\mathcal{L}(t)=|\langle \psi(t_0)|\exp(i\textrm{H}t/\hbar) \exp(-i\textrm{H}_{\varepsilon}t/\hbar)|\psi(t_0)\rangle|^2$, where $\textrm{H}_{\varepsilon}=\textrm{H} +\varepsilon V$ is a perturbed Hamiltonian on the original one $\textrm{H}$~\cite{Peres,Haug,Shres,Probst,WGWang04,WGWang05,Liu05}. For chaotic systems, the LE exponentially decays with time, with a rate proportional to the classical Lyapunov exponent, thereby demonstrating a kind of quantum-classical correspondence~\cite{WGWang02}. We numerically investigate the time evolution of the $\cal{L}$ for different $\eta$, setting $\textrm{H}_{\varepsilon}= p^2/2 + (g_{\varepsilon} + i\eta)|\psi(\theta)|^2\sum_{n}\delta (t-t_n)$ with $g_{\varepsilon}=g+{\varepsilon}$ and $\psi{(\theta,t_0)}=(\sigma/\pi)^{1/4} \exp (-\sigma \theta^{2}/2)$ with $\sigma=10$. Figure~\ref{LoschtEcho}(a) shows that for $\ehbar=4\pi$, the value of $1-\cal{L}$ increases super-exponentially with time when $\eta$ is nonzero. Note that, the $1-\cal{L}$ grows extremely quickly, so it exceeds the maximum double-precision value in a short time. In addition, the value of $1-\cal{L}$ for the quantum non-resonance case, i.e., $\ehbar=1$, also exhibits the super-exponential growth with time [see Fig.~\ref{LoschtEcho}(b)]. Intrinsically, under quantum resonance condition, the ratio
$R_{\mathcal{L}}(t)=[1-\mathcal{L}(t+1)]/[1-\mathcal{L}(t)]$
for different $\eta$ follows a scaling law, given by
$\ln[R_{\mathcal{L}}(t+1)]=\alpha [R_{\mathcal{L}}(t)\ln[R_{\mathcal{L}}(t)]]^{\beta}$, with $\alpha\approx 0.48$ and $\beta\approx 0.8$.
For $\ehbar=1$, the dependence of $\ln[R_{\mathcal{L}}(t+1)]$ on $R_{\mathcal{L}}(t)\ln[R_{\mathcal{L}}(t)]$ fluctuates around this scaling law [see Fig.~\ref{LoschtEcho}(c)]. In constructing $D$, the infinitesimal translation of the initial state in angular coordinate space effectively serves as a perturbation to the system. As a result, the $\cal{L}$ should be essentially equivalent to $D$. Our finding highlights the generality of the existence of the super-exponential instability in the NGPM model.

In noninteracting systems, the distance between two quantum states of Hermitian chaotic systems, characterized by $D$, remains unchanged with time evolution. Consequently, a widely employed metric for assessing dynamical instability is $\cal{L}$. Our findings reveal that, by taking into account nonlinear interactions, both $D$ and $\cal{L}$ can be utilized to quantify chaoticity for not only Hermitian systems but also Non-Hermitian systems [see Table.~\ref{SchemTabl}]. This expands our understanding of fundamental issues in quantum chaos.
\begin{table}[htbp]
\begin{center}
\begin{tabular}{|m{1.95cm}<{\centering}|m{1.95cm}<{\centering}|m{1.95cm}<{\centering}|m{2.15cm}<{\centering}|}
\hline Measurements
 & H-NI systems $(\eta=0,g=0)$ & H-I systems $(\eta= 0,g\neq0)$& NH-I systems $(\eta\neq 0,g\neq0)$\\
\hline Distance: $D$ & NA & Works & Works \\
\hline LE: $\cal{L}$ & Works & Works & Works \\
\hline
\end{tabular}
\caption{Measurements of dynamical instability in Hermitian noninteracting (H-NI), Hermitian interacting (H-I), and non-Hermitian interacting (NH-I) unstable systems.}\label{SchemTabl}
\end{center}
\end{table}

\section{Theoretical analysis and some discussions}
\subsection{Theoretical analysis}
Consider two initial states, $|\psi(t_0)\rangle$ and $|\varphi(t_0)\rangle=T(-\varepsilon)|\psi(t_0)\rangle$, where in real space $\varphi(\theta,t_0)=\psi(\theta+\varepsilon,t_0)$. Based on Eq.\eqref{Evoopt}, these two states produce two different Floquet operators, $U[\varphi(t_0)]$ and $U[\psi(t_0)]$, related by
\begin{equation}\label{Evooptequi0}
U[\varphi(t_0)]= T^{\dagger}(\varepsilon)U[\psi(t_0)]T(\varepsilon)\;.
\end{equation}
Using this equation, the states after the first kick at $t=t_1$ have a mathematical equivalence
\begin{align}\label{FKiKStates1}
|\varphi(t_1)\rangle &=U[\varphi(t_0)]|\varphi(t_0)\rangle= T(-\varepsilon)|\psi(t_1)\rangle\;,
\end{align}
and the corresponding Floquet operators at $t=t_1$ are given by \begin{equation}\label{Evooptequi1}
U[\varphi(t_1)]= T^{\dagger}(\varepsilon)U[\psi(t_1)]T(\varepsilon)\;.
\end{equation}
The fidelity of the states at $t=t_1$ can be obtained by using Eq.\eqref{FKiKStates1},
\begin{equation}\label{Fidelity1}
\langle\varphi(t_1)|\psi(t_1)\rangle=\langle\psi(t_1)|T(-\varepsilon)|\psi(t_1)\rangle\;.
\end{equation}
Repeating this process for the second kick, one can obtain the expression of both the states and Floquet operators
\begin{eqnarray}
|\varphi(t_2)\rangle &=&T(-\varepsilon)|\psi(t_2)\rangle\;,\\
U[\varphi(t_2)]&=& T^{\dagger}(\varepsilon)U[\psi(t_2)]T(\varepsilon)\;,\\
\langle\varphi(t_2)|\psi(t_2)\rangle&=&\langle\psi(t_2)|T(-\varepsilon)|\psi(t_2)\rangle\;.
\end{eqnarray}
It is straightforward to derive the expression for the fidelity after $n$-th steps using,
\begin{equation}\label{FidelityN}
\langle\varphi(t_n)|\psi(t_n)\rangle=\langle\psi(t_n)|T(-\varepsilon)|\psi(t_n)\rangle\;.
\end{equation}
which shows that the fidelity equals the expectation of the translation operator acting on the state $|\psi(t_n)\rangle$. Substituting this relation into Eq.\eqref{Distance} yields the expression for the distance, \begin{equation}\label{Distance2}
D(t_n)= 1- |\langle\psi(t_n)|T(\varepsilon)|\psi(t_n)\rangle|^2\;.
\end{equation}
To eliminate the contribution of the norm, we rescale the distance at an arbitrary time as
\begin{equation}\label{Distance3}
D(t)= 1- |\langle\psi(t)|T(\varepsilon)|\psi(t)\rangle|^2/\mathcal{N}^2_{\psi}(t)\;.
\end{equation}
Taking Eq.~\eqref{RSOTC} into account, we can obtain the equivalence
\begin{equation}\label{DistFOTO}
D(t)=C(t)\;.
\end{equation}
It is known that the $D$ measures how fast of two nearby states depart, thus straightforwardly captures the dynamical instability in quantum chaotic systems. Meanwhile, the $\mathcal{F}_O$ quantifies the information scrambling in many-body quantum systems. Our theoretical prediction in Eq.~\eqref{DistFOTO} unveils clearly the underlying relation between quantum chaos and quantum information scrambling. The numerically confirmed equivalence between $D$ and $C$, as depicted in Fig.~\ref{DisOTOCs}(a), demonstrates the simultaneous presence of the super-exponential instability and super-exponential scrambling.

Using the Taylor expansion $T(\varepsilon)\approx 1- i\varepsilon p/\ehbar$ for very small $\varepsilon$, it is straightforward to derive the following relation: $\mathcal{F}_{O}(t)\approx 1-\left({\varepsilon}/{\ehbar}\right)^2\left[\langle p^2(t)\rangle-\langle p(t)\rangle^2\right]$, where we neglect terms of order larger than the square of $\varepsilon$. Then, the distance $D$ is apparently connected to the mean energy
\begin{equation}\label{Distance3}
D(t)\approx \left(\frac{\varepsilon}{\ehbar}\right)^2\left[\langle p^2(t)\rangle-\langle p(t)\rangle^2\right]\;.
\end{equation}
In the absence of a directed current, i.e., $\langle p(t)\rangle=0$, the $\mathcal{F}_{O}(t)$ simplifies to 1 minus a term proportional to the mean energy, i.e., $\mathcal{F}_{O}(t) \approx 1 - \left({\varepsilon}/{\ehbar}\right)^2 \langle p^2(t) \rangle$, and the distance $D$ is directly proportional to the mean kinetic energy, i.e., $D(t)\approx \left({\varepsilon}/{\ehbar}\right)^2\langle p^2(t)\rangle$. The reveals clearly the underlying relation between dynamical instability and quantum diffusion.
\begin{figure}[t]
\begin{center}
\includegraphics[width=7.5cm]{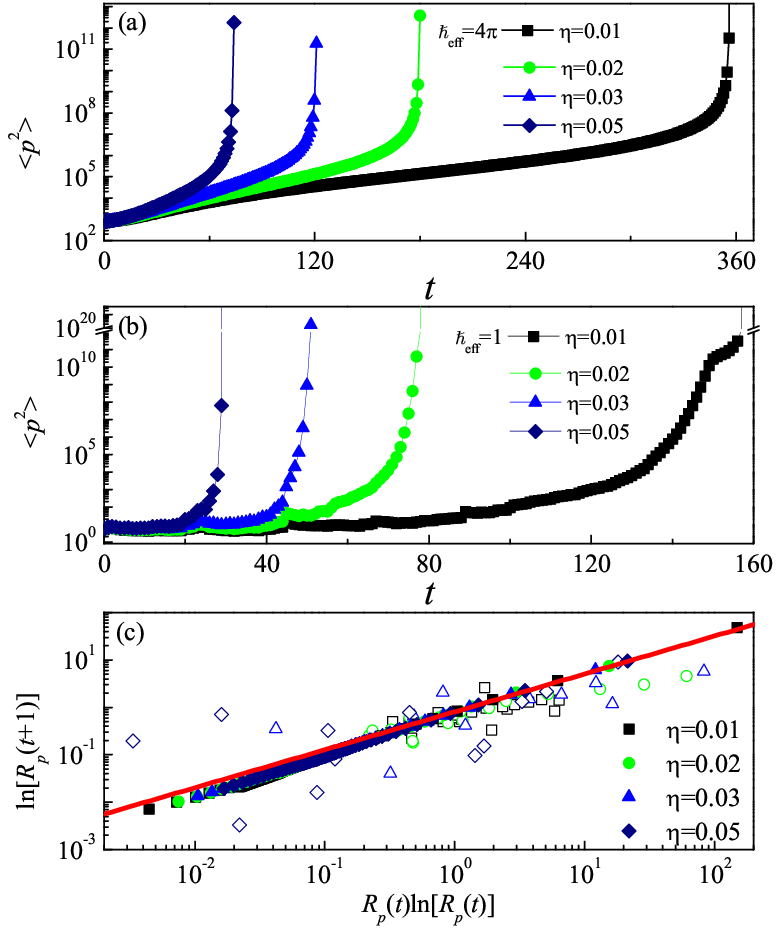}
\caption{(a) and (b): Time dependence of $\langle p^2\rangle$, in semi-log scale, for $\ehbar=4\pi$ in (a) and 1 in (b), with $\eta=0.01$ (squares), $0.02$ (circles), 0.03 (triangles), and 0.05 (diamonds). (c) The $\ln[R_{p}(t+1)]$ versus $R_{p}(t)\ln[R_{p}(t)]$, in log-log scale, for $\ehbar=4\pi$ (solid symbols) and 1 (empty symbols), with $\eta=0.01$ (squares), $0.02$ (circles), 0.03 (triangles), and 0.05 (diamonds). Solid line indicates the fitting function $\ln[R_{p}(t+1)]=\alpha [R_{p}(t)\ln[R_{p}(t)]]^{\beta}$, with $\alpha\approx 0.8$ and $\beta\approx 0.8$. Other parameters are the same as in Fig.~\ref{DisOTOCs}(a).}\label{FTQDiff}
\end{center}
\end{figure}

We further investigate numerically the time evolution of the mean energy for different values of $\eta$. Under the quantum resonance condition, $\langle p^2(t) \rangle$ exhibits a super-exponential increase over time, and larger values of $\eta$ lead to a faster growth of $\langle p^2(t) \rangle$ grows [See Fig.~\ref{FTQDiff}(a)]. Such super-exponential behaviors of quantum diffusion also emerges in quantum non-resonance condition [see Fig.~\ref{FTQDiff}(b). Interestingly, for $\ehbar=4\pi$, the ratio $R_{p}(t)=\langle p^2(t+1) \rangle/\langle p^2(t) \rangle$ for different $\eta$ is governed by a scaling law, i.e., $\ln[R_{p}(t+1)]=\alpha [R_{p}(t)\ln[R_{p}(t)]]^{\beta}$, with $\alpha\approx 0.8$ and $\beta\approx 0.8$. For $\ehbar=1$, the dependence of $\ln[R_{p}(t+1)]$ on $R_{p}(t)\ln[R_{p}(t)]$ exhibits some fluctuations around this scaling law [see Fig.~\ref{FTQDiff}(c)], indicating an essential feature of our system.
Therefore, the scaling law of energy diffuion determines the super-exponential behaviors both the dynamical instability and quantum scrambling. Note that the above found super-exponential growth of the mean energy is distinct from that observed in our previous work, i.e., Ref.~\cite{Zhao20}.

\begin{figure}[t]
\begin{center}
\includegraphics[width=7.5cm]{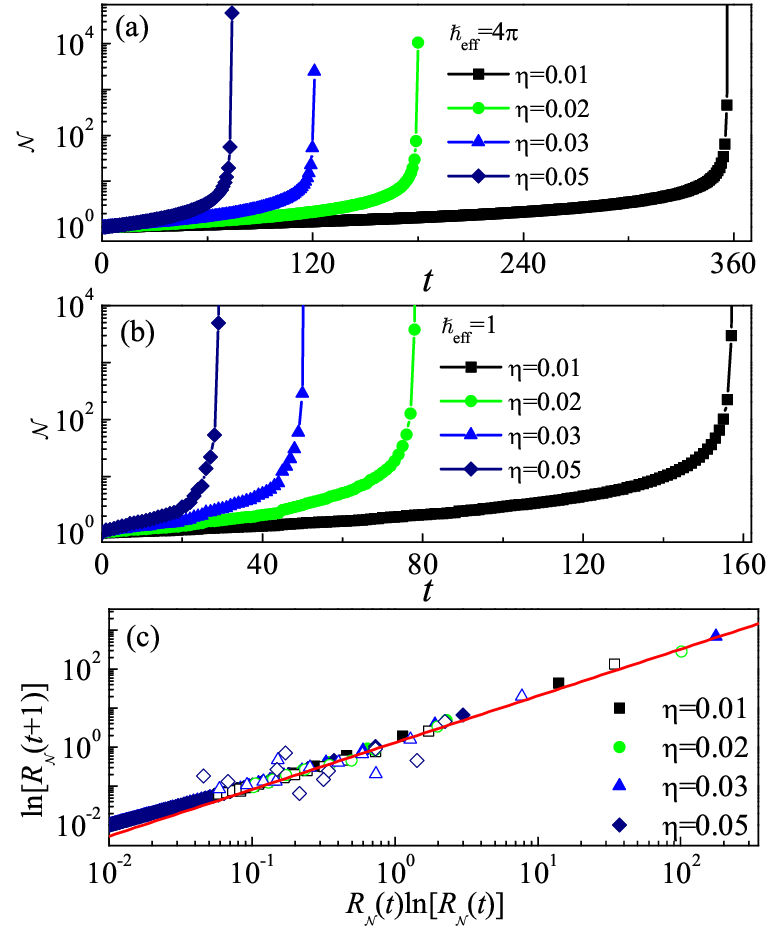}
\caption{(a) and (b): Time dependence of $\mathcal{N}$, in semi-log scale, for $\ehbar=4\pi$ in (a) and 1 in (b), with $\eta=0.01$ (squares), $0.02$ (circles), 0.03 (triangles), and 0.05 (diamonds). (c) The $\ln[R_{\mathcal{N}}(t+1)]$ versus $R_{\mathcal{N}}(t)\ln[R_{\mathcal{N}}(t)]$, in log-log scale, for $\ehbar=4\pi$ (solid symbols) and 1 (empty symbols), with $\eta=0.01$ (squares), $0.02$ (circles), 0.03 (triangles), and 0.05 (diamonds). Solid line indicates the fitting function $\ln[R_{\mathcal{N}}(t+1)]=\alpha [R_{\mathcal{N}}(t)\ln[R_{\mathcal{N}}(t)]]^{\beta}$, with $\alpha=1.3$ and $\beta=1.2$. Other parameters are the same as in Fig.~\ref{DisOTOCs}(a).}\label{FTNorm}
\end{center}
\end{figure}

It is widely accepted that the norm of a quantum state in non-Hermitian systems grows exponentially with time. Under the quantum resonance condition, however, the norm of a non-Hermitian kicked rotor exhibits super-exponential growth, leading to the the divergence of quantum diffusion and quantum scrambling~\cite{WLZhao24PRA}. We numerically investigate the time evolution of $\mathcal{N}=\langle \psi(t)|\psi(t)\rangle$ for different $\eta$. Our results demonstrate that for a specific value of $\eta$, the super-exponential increase of $\mathcal{N}$ occurs when the system is in both the quantum resonance ($\ehbar=4\pi$) and non-resonant ($\ehbar=1$) cases [see Figs.~\ref{FTNorm}(a) and (b)]. More importantly, for $\ehbar=4\pi$, the norm ratio $R_{\mathcal{N}}(t)=\mathcal{N}(t+1)/\mathcal{N}(t)$ for different $\eta$ obeys a scaling law given by
$\ln[R_{\mathcal{N}}(t+1)]=\alpha [R_{\mathcal{N}}(t)\ln[R_{\mathcal{N}}(t)]]^{\beta}$, with $\alpha=1.3$ and $\beta=1.2$. Under the quantum non-resonance condition, the dependence of $\ln[R_{\mathcal{N}}(t+1)]$ on $R_{\mathcal{N}}(t) \ln R_{\mathcal{N}}(t)$ also shows good agreement with this scaling law.

We analytically derive the scaling law of $R_{\mathcal{N}}(t)$ under the quantum resonance condition. Notably, obtaining an analytical solution for the scaling law in the quantum non-resonance case is challenging, as there is no explicit expression for the quantum state. For $\ehbar=4\pi$, the quantum state after successive kicks satisfies $\psi(\theta,t_{n+1})=\exp[-i(g+i\eta)|\psi(\theta,t_{n})|^2\ehbar/2\pi]\psi(\theta,t_{n})$, leading to the relation $|\psi(\theta,t_{n+1})|^2=\exp[\eta |\psi(\theta,t_{n})|^2/2\pi]|\psi(\theta,t_{n})|^2$. As a rough estimation, we approximate the probability density distribution using the norm, yielding
$\mathcal{N}(t+1)\propto \exp[\gamma \mathcal{N}(t)]\mathcal{N}(t)$, where $\gamma$ is a proportionality coefficient. It follows that $\ln[R_{\mathcal{N}}(t+1)]\propto R_{\mathcal{N}}(t)\ln[R_{\mathcal{N}}(t)]$, which is consistent with the scaling law shown in Fig~\ref{FTNorm}(c).

\subsection{Discussions and applications}

\subsubsection{Super-exponential growth of OTOCs in different forms}

\begin{figure}[b]
\begin{center}
\includegraphics[width=8.5cm]{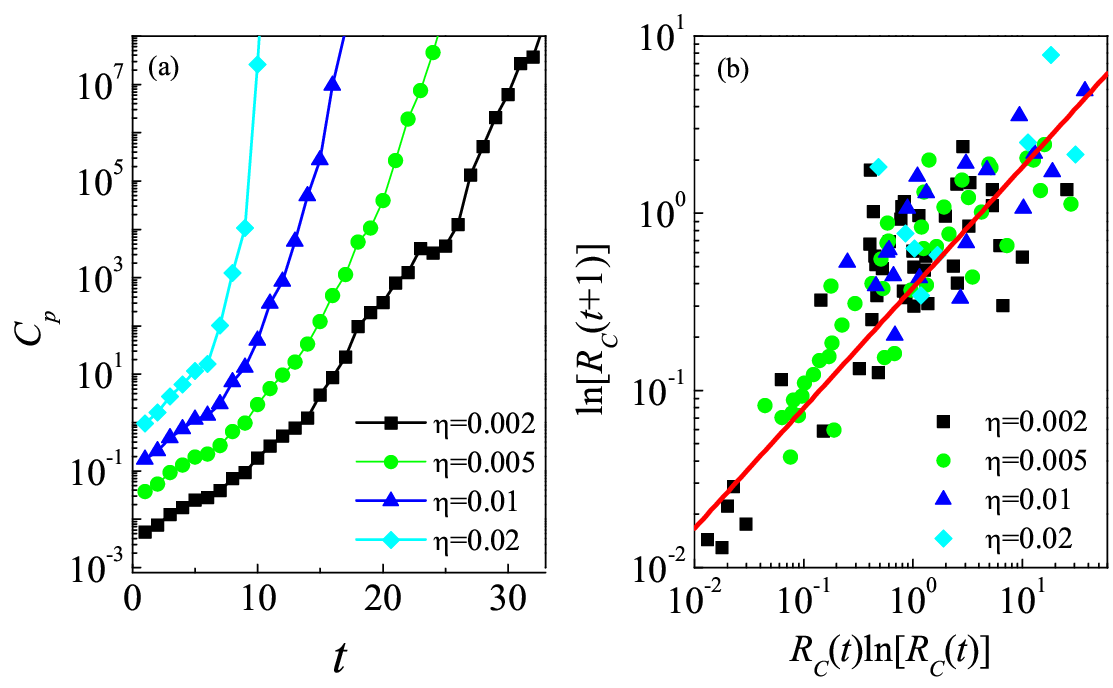}
\caption{(a) Time dependence of $C$, in semi-log scale, for $\varepsilon=10^{-5}$, with $\eta=0.002$ (squares), 0.005 (circles), $0.01$ (triangles), and 0.02 (diamonds). (b) The $\ln[R_{C}(t+1)]$ versus $R_{C}(t)\ln[R_{C}(t)]$, in log-log scale, for $\eta=0.002$ (squares), 0.005 (circles), $0.01$ (triangles), and 0.02 (diamonds). Solid line indicates the fitting function $\ln[R_{C}(t+1)]=\alpha [R_{C}(t)\ln[R_{C}(t)]]^{\beta}$, with $\alpha=0.38$ and $\beta=0.68$. The parameters are $g=0.5$ and $\ehbar=0.3$.}\label{NGPMAppen}
\end{center}
\end{figure}

The angular momentum operator has been commonly used to construct OTOCs in the delta-kicking systems to quantify the quantum scrambling~\cite{CYin21}. We thus investigate the OTOCs involving the angular momentum operator, in the form of $C=-\langle [W(t),V]^2\rangle$ with $W=e^{-i\varepsilon p/\ehbar}$ and $V=p$.
We numerically investigate the time evolution of $C$ under the quantum non-resonance condition. In our simulations, the initial state is a Gaussian function $\psi{(\theta,t_0)}=(\sigma/\pi)^{1/4} \exp (-\sigma \theta^{2}/2)$
with $\sigma=10$. Our results show that for a specific value of $\eta$, $C$ exhibits super-exponential increase over time [see Fig.~\ref{NGPMAppen}(a)]. We introduce the ratio $R_C(t)=C(t+1)/C(t)$ to characterize the super-exponential growth. Interestingly, the dependence of $\ln[R_C(t+1)]$ on $R_C(t)\ln[R_C(t)]$ approximately follows the relation  $\ln[R_{C}(t+1)]=\alpha [R_{C}(t)\ln[R_{C}(t)]]^{\beta}$, with $\alpha=0.38$ and $\beta=0.68$ [see Fig.~\ref{NGPMAppen}(b)], indicating the presence of a scaling law. Note that the time dependence of OTOCs in Ref.~\cite{WLZhao21} is not in similar law with respect to the super-exponential growth of our current work.

\subsubsection{NGPM for many-body Bosonic and optical fiber systems}

We consider Bose-Einstein condensates (BECs) in a ring trap~\cite{Kavoulakis03,JLiu06pra,CWZhang06prl} without an external potential and with a contact potential in dimensionless units, expressed as $V(r_2 - r_1) = (g + i\eta)\delta(r_2 - r_1)\sum_n\delta(t - t_n)$, where the delta modulation can be achieved using the Feshbach resonance technique. The imaginary part of the interaction describes the gain or loss of particles in the BECs~\cite{Keeling08,Sabari22}. The many-body Hamiltonian has the expression
\begin{equation}\label{MBHamil}
\text{H}=\int dr \hat{\Psi}^{\dagger}(r)\left[\frac{p^2}{2} + \frac{1}{2}g\hat{\Psi}^{\dagger}(r)\hat{\Psi}(r)\right] \hat{\Psi}(r)\;,
\end{equation}
where $\hat{\Psi}^{\dagger}(r)$ and $\hat{\Psi}(r)$ are creation and annihilation operators for Bosons at position $r$~\cite{Castin97}. Considering the quantum fluctuation of the uncondensed atoms, the annihilation operator can be decomposed as $\hat{\Psi}(r,t)=\Phi_{ex}(r,t)\hat{a}_{ex}(t) + \delta \hat{\Psi}(r,t)$, where the $\Phi_{ex}(r,t)$ denotes the condensed state and $\hat{a}_{ex}(t)$ is the corresponding annihilation operator, $\delta \hat{\Psi}(r,t)$ indicates the state of uncondensed atoms. There is an expansion of the condensed state $\Phi_{ex}(r,t)\approx \Phi(r,t)+ \Phi^{1}(r,t)/\sqrt{N}$, where the orders of larger than two has been neglected as $N\gg 1$. Taking this state to Eq.~\eqref{MBHamil} yields the mean-field approximation of the Hamiltonian
GP equation in Eq.~\eqref{NGPHamil}. The above approximate equivalence between the NGPM model and the second-quantized Hamiltonian of a BEC system provides a foundation for potentially applying our findings to a realistic many-body quantum system.

It is interesting to compare the behavior of OTOCs in the NGPM model and that in the second-quantized model. Actually, in previous works, the comparison between the OTOCs behavior in the second-quantized model and its semiclassical counterpart has been made for a general many-body two-mode system~\cite{ZYLi2022,Liu05,GFWang06pra,Jliu06pla}. The investigations on the Lipkin-Meshkov-Glick model of an interacting $N$-body two-mode system have revealed that the FOTOC exhibits exponential growth with time, indicating exponentially fast scrambling~\cite{ZYLi2022}. This exponential behavior corresponds to the classical chaos in mean-field phase space and can be also characterized by the exponential decay of $\mathcal{L}$~\cite{Liu05}. The mean-field description of the many-body two-mode system yields chaotic dynamics, such as the chaotic diffusion of classical trajectories in phase space~\cite{Jliu06pla,GFWang06pra}, implying the existence of exponential instability in the dynamics of wavepackets. Even though the above discussions are toward Hermitian systems, it is expected that the above correspondences still hold for the non-Hermitian systems as discussed in present work. Related investigations are ongoing.

On the other hand, optical fibers featuring a periodically modulated refractive index along the longitudinal direction have been employed to implement the quantum kicked rotor model, as reported in~\cite{Prange89,Agam92}. Considering that the refractive index is complex, optical settings provide an ideal platform for realizing systems that combine nonlinearity and non-Hermiticity~\cite{Ozdemir19}. We propose an optical setup to emulate the wave dynamics of the NGPM model, as described in Eq.~\eqref{NGPHamil}, where a series of periodically arranged multilayers of Kerr media in the propagation direction act as the temporally-modulated nonlinearity. The delta-kicking modulation of nonlinearity is achieved by setting the width of the Kerr media to be significantly smaller than the period of the optical sequence in the propagation direction~\cite{Fischer00,Rosen00,Prange89,Agam92}. Consequently, the propagation equation of the light is governed by the Hamiltonian in Eq.~\eqref{NGPHamil}. The super-exponentially fast energy diffusion of the NGPM model can be observed in the spatial frequency domain~\cite{Prange89,Agam92}, highlighting the emergence of both the super-exponential instability and super-exponential scrambling.

\subsubsection{Relations to quantum metrology}

The precision of a measurement is quantified by the minimal achievable uncertainty, i.e., the variance of an estimator $(\Delta \xi)^2$. The quantum-Cram{\' e}r-Rao bound predicts an inverse relationship between $(\Delta \xi)^2$ and quantum Fisher information (QFI) $I_Q$, specifically $(\Delta \xi)^2=1/I_Q$. It indicates that increasing $I_Q$ might improve the bound on the possible precision for the parameter $\xi$, if the right measurement can be made in an experiment, such as using quantum squeezing and entanglement~\cite{Giovannetti00}. Both theoretical and experimental investigations have reported the improvement of measurement precision of magnetic fields at least one order of magnitude  by using the Floquet driven potential~\cite{MJiang22prl}. Interestingly, delta-kicking potential can significantly increase $I_Q$, thereby enhancing the precision of magnetic field measurements~\cite{Fiderer18}. It has been proven that the QFI is proportional to LE, i.e., $I_Q = (1-\mathcal{F})/\xi^2$ for $\xi \ll 1$, where the LE is defined as $\mathcal{F}=\langle \psi(t_0)|U^{\dagger}(t)e^{-i\xi G}|U(t)\psi(t_0)\rangle$ with $G$ a Hermitian operator~\cite{Macri16,Fiderer18}. It is evident that $\mathcal{F}$ is equivalent to the FOTOC $\mathcal{F}_O$, obtained by substituting $G$ with $p$ and $\xi$ with $\varepsilon/\hbar$. In condition that $\varepsilon/\hbar \ll 1$, one can get the relation $I_Q=(\Delta p)^2$. Therefore, the superexponentally-fast growth of $(\Delta p)^2$ has potential application in improving the quantum metrology~\cite{ZYLi2022} [see Table.~\ref{ConcTabl}].

\begin{table}[htbp]
\begin{center}
\begin{tabular}{|m{2.05cm}<{\centering}|m{1.95cm}<{\centering}|m{3.2cm}<{\centering}|}
\hline
Quantity & Physical Meaning & Relation\\\hline
FOTOC ($\mathcal{F}_O)$ & Scrambling & $\mathcal{F}_O\propto \mathcal{L}_E\propto (\Delta p)^2$\\\cline{1-2}
LE ($\mathcal{L}_E$) & Instability & $I_Q\propto (\Delta p)^2$\\\cline{1-2}
QFI ($I_Q$) & Metrology & $(\Delta \xi)^2\propto\frac{1}{I_Q}\propto \frac{1}{(\Delta p)^2}$\\\cline{1-2}
Variance $(\Delta p)^2$ & Diffusion & The smaller $(\Delta \xi)^2$  indicates the higher precision measurement. \\\cline{1-2}
\hline
\end{tabular}
\caption{Relation of $\mathcal{F}_O$, $\mathcal{L}_E$, $(\Delta p)^2$, and $I_Q$, and their applications in quantum information scrambling, quantum instability, quantum metrology, and quantum diffusion.}\label{ConcTabl}
\end{center}
\end{table}

\section{Conclusion}
In this work, we uncover the super-exponential laws governing both the OTOCs $C$ and the distance $D$ (or LE) of quantum states, while also revealing their underlying connection with the super-exponentially fast diffusion of mean energy $\langle p^2\rangle$. Intrinsically, the super-exponential behaviors of these observables follow scaling laws that are independent of the non-Hermitian parameter $\eta$ as well as the effective Planck constant $\ehbar$, highlighting the significance of super-exponential instability, scrambling, and diffusion. These findings are realizable via the state-of-art experiments of BEC and optical systems. Our investigations also embrace the fundamental concepts of quantum scrambling, quantum instability, quantum diffusion~\cite{Lewis-Swan19}, and quantum metrology~\cite{Macri16,Linnemann16,Pezze18,ZYLi2022,Fiderer18}.

\section*{Acknowledgements}
W. Zhao is supported by the National Natural Science Foundation of China (Grant No. 12365002 and 12065009), the Science and Technology Planning Project of Jiangxi province (Grant No. 20224ACB201006 and 20224BAB201023),
Key Laboratory of Low Dimensional Quantum Materials and Sensor Devices of Jiangxi Education Institutes (NO. GanJiaoKeZi-20241301). J. Liu is supported by the NSAF (Contract No. U2330401).


\end{document}